\documentstyle[12pt]{article}
\title{
\vspace{2cm}
Relativistic QFTH-Couplings on the Worldline
\footnote{Talk presented by M.G.Schmidt at the Dubna Joint Meeting:
International Seminar on Path Integrals: Theory and Applications,
and 5th International Conference on Path Integrals from meV to MeV, 
Dubna, Russia, 27-31 May 1996}
}

\author{\large Michael G. Schmidt$^{a)}$, Christian Schubert$^{b)}$\\[3mm]
\em  $^{a)}$ Institut f\"ur Theoretische Physik,
     Universit\"at Heidelberg,\\
\em  Philosophenweg 16,
     D-69120 Heidelberg, Germany\\
\em  $^{b)}$ Institut f\"ur Theoretische Physik,
     Humboldt-Universit\"at zu Berlin,\\
\em  Invalidenstr. 110, D-10115 Berlin, Germany 
}
\date{}

\begin{document}

\maketitle

\begin{abstract}
In the framework of the worldline path integral approach 
to QFTH we discuss spin and relativistic couplings, 
in particular Yukawa and axial couplings to spin $\frac{1}{2}$,
and the case of spin 1 in the loop.
\end{abstract}

\vspace{-12cm}
\begin{flushright}
HD-THEP-96/43\\
HUB-EP-96/28
\end{flushright}

\thispagestyle{empty}
\newpage
\setcounter{page}{1}

{\parindent0em \bf Introduction}\\

Interacting relativistic particles are usually described by relativistic
local quantum \underbar{field} theory (QFTH); fields in
space-time $\phi_i(x^\mu)$ are subjected to canonical quantization
or they are varied in a path integral. This leads to the well-known
Feynman rules. However, already in the early days of QFTH there
was
a different approach \cite{1}: one considers the quantization of
relativistic \underbar{particles} moving on a space-time worldline
$x^\mu(\tau)$ and interacting with background fields.
We here restrict our discussion to effective 1-loop actions, i.e.
the particles are running on a closed loop.

For a massive complex scalar in the loop one has (euclidean action;
singularity at $T=0$ to be renormalized)
\begin{eqnarray}\label{1}
\Gamma_{\rm scalar}&=&-\log\det(-{\cal D}^2+{\bf V}
(x)+m^2)
\nonumber\\
&=&\int^\infty_0\frac{dT}{T} e^{-m^2T}\int_{x(T)=x(0)}
[Dx(\tau)]{\rm Tr} P\exp
\left\{-\int^T_0
{\cal L}^B_{WL}d\tau\right\}\end{eqnarray}
where the path integral describes quantum mechanics
and the Schwinger
$T$-integral takes care of relativity. The worldline Lagrangian is
\begin{equation}\label{2}
{\cal L}^B_{WL}=\frac{\dot x^2_\mu}{4}+ig\dot x_\mu
(\tau){\bf A}_\mu(x(\tau))+{\bf V}(\phi(x(\tau))\end{equation}
if one has an (possibly nonabelian, matrix valued in color space)
gauge background field and a ``potential'' (the second derivative
of the QFTH-potential), e.g. $\sim\phi,\phi^2$ for $\phi^3,\phi^4$
theory
resp. Indeed this ``first quantization'' and the background
field method were used intensively in modern string theory. In
particular
the recurrence to relativistic particles helped to write down
nonlinear
$\sigma$-models for $x^\mu(\sigma,\tau)$ on the ``world sheet''
\cite{2}.
In the limit string tension $\frac{1}{\alpha'}\to\infty$ one should
obtain
the well-known local QFTH's \cite{4}. This was studied carefully by Bern
and
Kosower \cite{3} for tree and 1-loop amplitudes. The resulting
rules -- equivalent to  but looking much different from
the Feynman
rules --
are very effective in particular if combined with
the spinor
helicity formalism in QCD. 

Later on Strassler \cite{5} argued that
these
rules can partially be obtained from worldline path
integrals.
The most important ingredient in the worldline approach is the
notion
of worldline Green functions (GF). For the circle  a convenient
GF \cite{5,6} corresponding to the free part ${\cal L}_0=\dot
x^2_\mu/4$
is\\
\begin{equation}\label{3}
G_B(\tau,\tau')=|\tau-\tau'|-\frac{(\tau-\tau')^2}{T}.\end{equation}
It fulfils\\
\begin{equation}\label{4}
\frac{1}{2}\frac{\partial^2}{\partial\tau^2} G_B(\tau,\tau')
=\delta(\tau-\tau')-\frac{1}{T}\end{equation}
i.e. a constant zero mode $x^{(0)}_\mu$ is taken out of
$x_\mu(\tau)=x^{(0)}
_\mu+y_\mu(\tau)$ and is fixed by the orthogonality $y\bot
x^{(0)}$ leading
to $\int^T_0 d\tau y_\mu(\tau)=0$. Thus $x^{(0)}$ is the ``center
of mass''-
coordinate
of the loop. More general background charges $\rho(\tau)$ instead of
$\frac{1}{T}$ in (\ref{4}) are physically equivalent but 
less convenient because they break
the translation invariance in $\tau$. 

Splitting $\int[Dx]$ as
$\int dx^{(0)}\int[Dy]$ we can evaluate worldline path integrals 
contracting 

\begin{equation}\label{5}
\langle y_\mu(\tau) y_\nu(\tau')\rangle=-\delta_{\mu\nu} G_B
(\tau,\tau').\end{equation}

\noindent
as we are used from QFTH.
This allows to evaluate 1-loop  effective
actions
very efficiently. In the Fock-Schwinger gauge 

$${\bf
A}_\mu(x^{(0)}+y)
=y^\rho\int^1_0 d\eta\eta{\bf F}_{\rho\mu}(x^{(0)}+\eta y)$$

\noindent
 and with
covariant
Taylor expansion of ${\bf F}_{\rho\mu}$ and ${\bf V}$ 
at $x^{(0)}$ one obtains, for
scalar
loops, \cite{7}
\begin{eqnarray}\label{6}
\!\!\!\!\!\!\!&&\Gamma[{\bf F},{\bf V}]=\int dx^{(0)}
\int^\infty_0\frac{dT}{T}e^{-
m^2T}\sum^\infty_{n=0}\frac{(-1)^n}{n}\int[Dy]{\rm tr}_
{\rm color}P\exp
\left\{-\int^T_0 d\tau\frac{\dot y^2}{4}\right\}\\
\!\!\!\!\!\!\!
&&\times
\int\limits^{\tau_1=T}_0 \!\!\!d\tau_2\ldots\!
\int\limits^{\tau_{n-1}}_0
\!\!\!d\tau_n\prod^n_{j=1}\left[e^{y(\tau_j)\cdot D}{\bf V}(x^{(0)})+ig
\dot y^{\mu_j}
(\tau_j)y^{\rho_j}(\tau_j)
\int^1_0 d\eta_j\eta_j e^{\eta_j y(\tau_j)\cdot D}
{\bf F}_{\rho_j\mu_j}
(x^{(0)})\right].\nonumber\end{eqnarray}
In the case where only the potential 
is present this immediately leads to
the
minimal number of invariants. Also for the case of gauge
interactions it
is the most efficient form for obtaining the heat kernel expansion.
Such
expansions can be used in the calculation of fluctuation corrections
around
quasiclassical configurations.

The effective action (euclidean notation) for gauge fields ${\bf
A}_\mu$
coupling minimally to Dirac particles ${\cal L}=\bar\psi{\bf O}\psi$
with
${\bf O}={\bf O}^+=(\partial_\mu+ig{\bf A}_\mu)\gamma_\mu$ is
\begin{equation}\label{7}
\Gamma_{\rm Dirac}=\log\det{\bf O}=\frac{1}{2}{\rm Tr}\log{\bf
O}
{\bf O}^+=\frac{1}{2}{\rm Tr}\log(-D^2_\mu{\bf
1}+g\sigma_{\mu\nu}
{\bf F}_{\mu\nu}).\end{equation}
This contains a Dirac matrix valued potential but otherwise is
similar
to the scalar theory (``second order formalism''). One can
substitute
the Dirac matrix trace
by a 
Grassmann variable integration over $\psi(\tau)$
\cite{6,5,9}
\begin{equation}\label{8}
\Gamma_{\rm Dirac}=-
2\int^\infty_0\frac{dT}{T}\int_{x(0)=x(T)}[Dx]\int_{\psi(0)=-
\psi(T)}[D\psi] {\rm tr} P \exp\left\{-\int^1_0
L^D_{WL}d\tau\right\}\end{equation}
where ${\cal L}^D_{WL}={\cal L}^B_{WL}+{\cal L}^F_{WL}$ now
contains
besides the bosonic piece (\ref{2}) a fermionic part
\begin{equation}\label{9}
{\cal L}^F_{WL}=\frac{1}{2}\psi_\mu\dot\psi_\mu(\tau)-ig
\psi_\mu(\tau)\psi_\nu(\tau)
{\bf F}_{\mu\nu}(x(\tau))\end{equation}
with a corresponding Green function\\
\begin{equation}\label{10}
\langle\psi_\mu(\tau)\psi_\nu(\tau')\rangle=\frac{1}{2}\delta_{\mu\nu}
G_F(\tau,\tau')=\frac{1}{2}\delta_{\mu\nu}{\rm sign}(\tau-
\tau').\end{equation}
Surprisingly ${\cal L}^D_{WL}$ is supersymmetric (a remnant of
local SUSY),
i.e. invariant under
$\delta x_\mu=- 2\eta\psi_\mu,\delta\psi_\mu=\eta
\dot x_\mu$
with a constant
Grassmann parameter $\eta$,
though the boundary conditions break the SUSY.
Introducing
superfields in superspace $\hat\tau(\tau,\theta)(\int
d\theta\theta=1):$
$X_\mu(\hat\tau)=x_\mu(\tau)+\sqrt2\theta\psi_\mu(\tau)$
and a superderivative
$D=\frac{\partial}{\partial\theta}-
\theta\frac{\partial}
{\partial\tau}$
the worldline action can be written as
\begin{equation}\label{14}
\int^T_0d\hat\tau{\cal L}^D_{WL}=\int^T_0d\tau\int d\theta
\hat{\cal L}^D_{WL}
=\int
d\hat\tau\left(\frac{1}{4} DX_\mu D^2X_\mu-ig DX_\mu
{\bf A}_\mu(X)\right).\end{equation}
Thus the formalism is manifestly supersymmetric
\cite{6,8,10} and one can get
the
effective action with Dirac particles in the loop from the one with
scalars by just substituting super Green functions\\
\begin{equation}\label{15}
\hat G(\hat\tau,\hat\tau')=G_B(\tau,\tau')+\theta\theta'
G_F(\tau,\tau')
\end{equation}
in the bosonic calculation.
This also explains the so-called
``chain''-rule: substitute (index-) closed chains of $\dot
G_B(\tau,\tau')
=\frac{\partial}
{\partial\tau} G_B(\tau,\tau')={\rm sign}
(\tau-\tau')-2(\tau-\tau')/T$ by the same chain minus the
corresponding
chain of $G_F$'s in the Dirac case. Note that SUSY unbroken by the
b.c.'s would imply $\dot G_B=G_F$ and hence a vanishing of such
terms. 
\vskip10pt

{\parindent0em \bf Further spin $\frac{1}{2}$ couplings}\\

Curiously the Yukawa coupling ${\cal L}^{\rm Yuk}=\bar\psi(-i \lambda
\phi)\psi$ was not translated to the worldline formulation until
our recent work with M. Mondrag\'on and L. Nellen
\cite{11}. In a dimensional reduction approach it can be interpreted as a gauge coupling
in a fifth dimension (after some redefinition of $\gamma_\mu$).
Introducing a new superfield $X_5=\tilde x_5+\sqrt2\theta\psi_5$ or its more
convenient  superderivative $(x_5\equiv-\dot{\tilde x}_5)$
\begin{equation}\label{16}
\bar X=\sqrt2 DX_5=\sqrt2\psi_5+\theta x_5
\end{equation}
the (super) worldline Lagrangian is extended by
\begin{eqnarray}\label{17}
\int d\theta\hat{\cal L}^{\rm Yuk}_{WL}&=&\int d\theta
\left(\frac{1}{4}\bar X D\bar X+i\lambda\phi
(X_\mu)\bar X(\hat\tau)\right)\nonumber\\
&=&\frac{x^2_5}{4}+\frac{1}{2}\psi_5\dot\psi_5+i\lambda
\left(x_5\phi(x)-
2\psi_5\psi_\mu\partial_\mu\phi(x)\right).\end{eqnarray}
Integrating out the auxiliary field $x_5$ we obtain
\begin{equation}\label{18}
\frac{1}{2}\psi_5\dot\psi_5+\lambda^2\phi^2-2i\lambda\psi_5\psi_\mu
\partial_\mu\phi(x).\end{equation}
Substituting  $\phi$ by $m/\lambda$ gives the
well-known mass term \cite{12}.

For a more systematic treatment \cite{13} consider now 
in analogy
to eq. (\ref{7}) a Dirac operator (euclidean)
\begin{equation}\label{19}
{\bf O}=(\partial_\mu+ig V_\mu+ig\gamma_5 A_\mu)\gamma_\mu-im
-i\lambda\phi+\gamma_5\lambda'\phi'\end{equation}
with scalar $(\phi)$, pseudoscalar $(\phi')$, vector $(V_\mu)$ and
axial vector $(A_\mu)$ couplings. The operator ${\bf O}^+$ has a
change in sign in the $A_\mu$ and $m,\phi$ terms. Thus we substitute
(\ref{7}) by\\
\begin{equation}\label{20}
\Gamma={\rm Tr}\log{\bf O}=\frac{1}{2}{\rm Tr}(\log {\bf O}
{\bf O}^+)+\frac{1}{2}{\rm Tr}(\log {\bf O}-\log{\bf O}^+)
\end{equation}
with
\begin{eqnarray}\label{21}
{\bf O}{\bf O}^+&=&-D^2+g\sigma_{\mu\nu} V_{\mu\nu}+g_5
\gamma_5\sigma_{\mu\nu} A_{\mu\nu}+i\lambda\gamma_\mu\partial_\mu\phi
\nonumber\\
&&-\lambda'\gamma_5\gamma_\mu\partial_\mu\phi'+2(im+i\lambda\phi
-\gamma_5\lambda'\phi')ig_5\gamma_5\gamma_\mu A_\mu+m^2\nonumber\\
&&+2m\lambda\phi+\lambda^2\phi^2+\lambda'^2\phi'^2\end{eqnarray}
where $V_{\mu\nu},A_{\mu\nu}$ are the field strengths.
The derivative of the second term in (\ref{20}) with respect to
the
background field $U (U=A,\phi'$ in the following) can be written as
\begin{equation}\label{22}
\frac{\delta}{\delta U}{\rm Tr}(\log {\bf O}-\log {\bf O}+)=
{\rm Tr}\left(\left\{\frac{\delta{\bf O}}{\delta U}{\bf O}^+-
{\bf O}\frac{\delta{\bf O}^+}{\delta U}\right\}\frac{1}{{\bf O}
{\bf O}^+}\right).\end{equation}
This ``second order'' formalism can be translated to the worldline
formulation with a purely Grassmann even Lagrangian if one introduces two
additional superfields $\bar X$ and $X'$ appearing in the
coupling to scalars / pseudoscalars as in (\ref{17}).
The new piece of the worldline action is particularly simple if
expressed in superfields. We proposed \cite{13}\\
\begin{eqnarray}\label{23}
S_{WL}&=&\int d\tau d\theta\biggl[
{1\over 4}DX_{\mu}D^2X_{\mu}
+{1\over 4}\bar XD\bar X
+{1\over 4} X'D X'
\\
&&
+i\lambda\phi(X)\bar X
+i{\lambda}'{\phi}'(X)X'
-igDX_{\mu}V_{\mu}(X)
+g_5\bar X X'DX_{\mu}A_{\mu}(X)
\biggr]\nonumber
\end{eqnarray}
with the original
worldline supereinbein field $\Lambda=e+\sqrt{e}\theta
\chi$ gauge
fixed to $e=2,\ \chi=0$ on the circle. The form of the
axial coupling with both $\bar X,X'$ involved is surprising. Written out in
components
this is a lengthy expression \cite{13}.

The imaginary part (\ref{22}) can be worked out 
and also translated to worldline
language
\begin{equation}\label{24}
\Gamma'_U=\frac{\delta}{\delta U} i{\rm Im}\Gamma=-2
\int^\infty_0\frac{dT}{T}\int[DX_\mu][D\bar X][DX'](-1)^F\Omega_U
e^{-S_{WL}}\end{equation}
with\\
\begin{equation}\label{25}
\Omega_{\phi'}=-i\lambda'\sqrt{2}
\int d\tau d\theta
\theta(-{1\over 4}D^2X_\mu
DX_\mu+{1\over 4}
\bar X D\bar X)X'\end{equation}
and a similar expression (integrated product of two superoperators
appearing in $S_{WL}$!) for $\Omega_{A_\mu}$. (\ref{24})
besides the action term
 contains the Witten index operator $(-1)^F$ changing the fermionic
b.c.'s from antiperiodic to periodic, and the $\Omega_U$. The
$\Gamma_{U}'$ have to be integrated over the field $U$ for getting the
action (An alternative way was given in \cite{14}.). Interestingly
$(-1)^F$ plays the role of $\gamma_5$, a well-known fact in the
discussion of anomalies \cite{15}. The change to periodic b.c.'s
allows for constant Grassmann modes $\psi_0$. Grassmann integration
then requires each of the modes to be present once. This leads to
$\epsilon$-tensors. 

We have tested our worldline action by
calculating \cite{13} a collection of amplitudes first in the conventional
Dirac-Feynman, then in second order, and finally in 
the worldline formalism.
In particular, we could reproduce PCAC and the axial anomaly;
we can also derive higher order terms in the 1-loop effective action.

In recent publications \cite{14,16} it was argued that ${\bf O}$ and
${\bf O}^+$ can be arranged in a $8\times 8$ spinor matrix
corresponding to a 6-dimensional space, thus explaining the need for
our fields $\bar X,X'$. The 6-dimensional $\gamma$-matrices
can be substituted by Grassmann $\psi$'s and our
worldline Lagrangian is reproduced and generalized in an elegant way.
There is a subtle technical
point \cite{13,17} about  either
integrating out the auxiliary fields
$x_5,x_6$ belonging to $\bar X,X'$ or contracting them in Green
functions if the correct field theory results are to be
obtained.

\vskip10pt

{\parindent0em \bf Spin 1 in the loop}\\

The case of massless spin 1 gauge bosons coupling to 
a gauge boson
background is particularly important in QCD applications. In the
covariant 't Hooft-Feynman background gauge the 1-loop action
has the Schwinger form
\begin{equation}\label{26}
\Gamma(A)=\frac{1}{2}\int^\infty_0\frac{dT}{T}{\rm Tr}\exp
(-\hat h T)\end{equation}
with the (color and Lorentz) matrix valued  Hamiltonian
\begin{equation}\label{27}
\hat h^{ab}_{\mu\nu}=-D^{ac}_\rho D^{cb}_\rho\delta_{\mu\nu}
-2ig{\bf F}_{\mu\nu}^{ab}(x).\end{equation}
One can evaluate this by a 
``bosonized'' worldline path integral including a matrix
valued potential but motivated by the spin $\frac{1}{2}$ case one
might try to substitute Lorentz (and eventually also color) indices
by Grassmann variables. Building on an earlier proposal
\cite{5} M. Reuter and the present authors
have shown \cite{18} that a worldline Hamiltonian
\begin{equation}\label{28}
\hat H_0=(\hat p_\mu+g A_\mu(\hat x))^2-:\hat{\bar\psi}_\mu
(2ig F^{ab}_{\mu\nu}) \hat\psi_\nu:\end{equation}
has the required properties. Here
$\hat p_\mu=-i\frac{\partial}{\partial x_\mu},\ \hat{\bar\psi}
=\frac{\partial}{\partial\psi_\mu}$, and $\bar H_0$ 
acts on forms $\phi(x,\psi)$
depending on $x$ and a set of classical Grassmann variables
$\psi_{\mu}$.
The
second term in (\ref{28}) needs  to be anti-Wick ordered.
$\hat H_0$ then acts on 1-forms (in $\psi$) just as the
operator (\ref{27}). Thus one has to project onto the one-forms.
This requires the introduction of a kind of mass term $C\hat\psi_\mu
\hat{\bar\psi}_\mu$ in $H_0$ with $C$ to be taken to infinity
at the end. Furthermore, a GSO type projector $(1-(-1)^F)$ has to be
applied. Besides this the anti-Wick ordering has to be
expressed in terms of a Weyl ordered $\hat H_{\rm Weyl}$
appropriate for transcription to a worldline 
path integral with
midpoint rule. Altogether we obtain an action
\begin{eqnarray}\label{29}
&&\Gamma(A)=-\frac{1}{2}\lim_{C\to\infty}\int^\infty_0\frac{dT}{T}
e^{-CT(D/2-1)}\int[Dx_\mu]\frac{1}{2}\left(
\int_{\rm antiper.}-\int_{\rm per.}\right)[D\psi_\mu]
[D\bar\psi_\mu]\nonumber\\
&&\times{\rm Tr} P \exp\left[-\int^T_0 d\tau
\left\{\frac{\dot x^2_\mu}{4}+ig\dot x_\mu A_\mu+\bar\psi_\mu
\left((\partial_\tau-C)\delta_{\mu\nu}-2ig {\bf F}_{\mu\nu}
\right)\psi_\nu\right\}\right]\end{eqnarray}
and fermionic Green functions
\begin{equation}\label{30}
G^C_{{\rm per.}\atop{\rm antiper.}} =-\left[\theta
(-\tau)\pm\theta(\tau) e^{-CT}\right] e^{C\tau}/(1\mp e^{-CT})
\end{equation}
similar but not identical to the expressions in \cite{5}. We should
stress that (\ref{29}) is the result of a rigorous derivation, will
lead to correct gluon amplitudes, and thus is a good starting point
for deriving Bern-Kosower type 1-loop rules. As a simple application
we have rederived \cite{18} the 1-loop action in a constant 
pseudo-abelian background.

Considering the multiloop case it is not so obvious that introducing
Grassmann variables to represent spin is an equally
 efficient way as it is
for spin $\frac{1}{2}$. A 2-loop derivation directly from
string theory would be very illuminating. For first steps into this
direction see \cite{19}. The multiloop formulation of QFTH in the
worldline language will be subject to another contribution in
these proceedings.

\newpage
\newcommand{\bib}[1]{\bibitem{#1}}

\end{document}